\documentclass[preprint2]{emulateapj}

\usepackage{graphicx}
\usepackage{longtable}
\usepackage{lineno}

\defcitealias{U3}{Paper~III}
\newcommand{\fer}{{\it Fermi}}
\newcommand{\wise}{{\it WISE}}
\newcommand{\paperiii}{{\citetalias{U3}}}

\slugcomment{version \today}
\shorttitle{Unidentified $\gamma$-ray Sources VI}
\shortauthors{M. Nori et al. 2013}

\begin{document}
\title{Unveiling the nature of the unidentified $\gamma$-ray sources VI: \\ $\gamma$-ray blazar candidates in the WISH survey and their radio properties}

\author{
  M. Nori\altaffilmark{1,2}, 
  M. Giroletti\altaffilmark{1},
  F. Massaro\altaffilmark{3}, 
  R. D'Abrusco\altaffilmark{4}, 
  A. Paggi\altaffilmark{4}, 
  G. Tosti\altaffilmark{5,6},
  S. Funk\altaffilmark{3}
}

\altaffiltext{1}{INAF Istituto di Radioastronomia, via Gobetti 101, 40129, Bologna, Italy}
\altaffiltext{2}{Dipartimento di Scienza Applicata e Tecnologia, Politecnico di Torino, Corso Duca degli Abruzzi 24, 10129, Torino, Italy}
\altaffiltext{3}{SLAC National Laboratory and Kavli Institute for Particle Astrophysics and Cosmology, 2575 Sand Hill Road, Menlo Park, CA 94025, USA}
\altaffiltext{4}{Harvard - Smithsonian Astrophysical Observatory, 60 Garden Street, Cambridge, MA 02138, USA}
\altaffiltext{5}{Dipartimento di Fisica, Universit\`a degli Studi di Perugia, 06123 Perugia, Italy}
\altaffiltext{6}{Istituto Nazionale di Fisica Nucleare, Sezione di Perugia, 06123 Perugia, Italy}

\begin{abstract}
According to the second \textit{Fermi} LAT Catalog (2FGL), about one third of the $\gamma$-ray sources listed have no assigned counterparts at lower energies. Many statistical methods have been developed to find proper counterparts for these sources.
We explore the sky area covered at low radio frequency by Westerbork in the Southern Hemisphere (WISH) survey to search for blazar-like associations among the unidentified $\gamma$-ray sources listed in the 2FGL (UGSs).  
Searching the WISH and NRAO VLA Sky Survey (NVSS) radio surveys within the positional uncertainty regions of the 2FGL UGSs, we select as $\gamma$-ray blazar candidates the radio sources characterized by flat radio spectra between 352 MHz and 1400 MHz. We propose new $\gamma$-ray blazar associations for eight UGSs and we also discuss their spectral properties at low radio frequencies.
 {We compare the radio flux density distribution of the low radio frequency $\gamma$-ray blazar candidates with that of $\gamma$-ray blazars associated with other methods. We find significant differences between these distributions.} Finally, we discuss the results of this association method and its possible applicability to other regions of the sky and future radio surveys.
\end{abstract}


%

\section{Introduction}

In the recent decades, $\gamma$-ray astrophysics has made great advances enabled by improvements in instrumentation in the high energy technologies. The most recent and most accurate $\gamma$-ray source catalog is the \textit{Fermi} Large Area Telescope (LAT) Second Source Catalog  \citep[2FGL,][]{2FGL}, based on the first 24 months of data from LAT.  Thanks to its silicon strip pair production and modern analysis processes, LAT has drastically reduced the positional error of the sources with respect to previous studies, like those performed by the Energetic Gamma-Ray Experiment Telescope (EGRET) on board the Compton Gamma-Ray Observatory \citep{CGRO}.

As the \fer-LAT $\gamma$-ray observations of the sky continue, the task of finding clear counterparts to all the detected sources becomes increasingly challenging, mostly due to the large positional uncertainty of the faintest sources. Since a strong connection between $\gamma$-ray and radio emission has been clearly demonstrated \citep{Ghirlanda2010,Mahony2010,Ackermann2011a}, it is natural to exploit radio surveys for finding new associations.
Several association methods have been proposed to match the $\gamma$-ray sources detected with source catalogs at lower frequencies \citep {U4,Masetti2013}, to give a proper counterpart to unidentified sources, where possible. Of the 1873 sources in the 2FGL catalog, 575 have unknown physical nature. Regarding the \textit{Fermi} sources of known types, blazars are the largest known population representing more than 80\%. It is, therefore, fair to presume that a significant fraction of the unidentified $\gamma$-ray sources (UGSs) can be unrecognized blazar-like sources. Finding them is the aim of our work. \par

Blazars are compact radio sources with flat (i.e., spectral index $\alpha< 0.5$, where $\rm{S}_\nu  \sim \nu^{-\alpha}$) radio spectra that steepen towards the infrared-optical bands; their overall spectral energy distribution shows two broadband components: the low energy one peaking in the infrared (IR) to X-ray frequency range, while the high energy one extends up to the MeV-to-TeV range. Their emission features high and variable polarization, apparent superluminal motions  of radio jet components and high apparent luminosity, coupled with rapid flux variability over the whole electromagnetic spectrum. Blazars come into two classes: flat-spectrum radio quasars and BL Lac objects, which we label here as BZQs and BZBs respectively, following Roma-BZCAT \citep{RomaBZCAT} nomenclature. Blazar emission interpreted as originating in radio loud active galaxies, with one of the relativistic jets emerging from the super-massive black hole in the galaxy nucleus pointing almost directly to the observer \citep{Giommi2012}. \par

This paper is the latest of a series that focus on the nature of UGSs: \citet{U1} (hereafter Paper I), which investigates sources as UGSs counterparts as described in \citet[][Paper II]{U2}; \citet[][Paper III]{U3} adds another feature for the search of blazar-like sources focusing on the low-frequency radio feature of blazars; \citet[][Paper IV]{U4} involves the X-ray emission as a distinctive feature and \citet[][Paper V]{U5} propose a renewed IR approach, based on a 2-dimensional kernel density estimation (KDE) technique, all for the same purpose.

In this work, we apply the method proposed in \paperiii\ to search for blazar-like candidates within the $\gamma$-ray positional uncertainty regions of the UGSs listed in the 2FGL and to select counterpart among them. This method is based on a multifrequency procedure, relying primarily on the flatness of the radio spectrum. The innovative point in this method is that it is based on low-frequency (i.e., $<1$ GHz) measurements, which is unprecedented in the investigation the emission of blazars and their association with UGSs. In \paperiii, we combined the observations at 325\,MHz from the Westerbork Northern Sky Survey (WENSS) at 325\,MHz with those of the NRAO Very Large Array Sky Survey (NVSS) at 1.4\,GHz.
In this work, we apply the same approach to the region covered by the counterpart of WENSS in the southern sky: the Westerbork in the Southern Hemisphere \citep[WISH,][]{De Breuck2002}.
Thanks to the combined results of this new search and of \paperiii, we build a reasonably sized population of new low-frequency selected blazar candidates. We study the spectral properties of individual candidates and explore the flux density distribution of the whole sample in comparison to the $\gamma$-ray blazars already listed in the 2nd \textit{Fermi} LAT Catalog of Active Galactic Nuclei \citep[2LAC,][]{2LAC}.  We further refine our search by looking for IR, optical, UV and X-ray counterparts within the catalogs available.

The paper is organized as follows: in Sect.\ 2 we describe our method; in Sect.\ 3 we report and characterize our list of candidates; in Sect.\ 4 we discuss the spectral properties of all the sources found with the low-frequency method and make some projections regarding expectations from new low-frequency radio telescopes and surveys. 

\section{Method}
\subsection{Blazar characteristics in the low-frequency radio band}

The characteristics of blazars (in particular $\gamma$-ray emitting blazars) in this range of the electromagnetic spectrum are discussed in \paperiii; we briefly summarize those results here for the sake of clarity. 

Based on the cross correlation between the 2FGL catalog, the Roma-BZCAT v4.1 (the most comprehensive blazar catalog up to now), and the WENSS, we defined one sample labeled as Low radio frequency Blazar (LB) and a subsample labeled Low radio frequency $\gamma$-ray Blazar (LGB).
Since the Roma-BZCAT catalog is based on the NVSS survey, we computed the low-frequency radio spectral index values for these samples as:            

\begin{equation}
\centering
\label{eq:index}
\alpha_{\nu}^{1400}=\log{ \left( \frac{S_{1400}}{S_{\nu}} \right )}/\log{\left( \frac{\nu}{1400} \right)}
\end{equation}

{where $\nu$ is the low radio frequency measured in MHz (i.e., 325 MHz for WENSS and 352 MHz for WISH), $S_{1400}$ is the flux density measured at 1400 MHz in the NVSS survey and $S_\nu$ is the flux density measured at $\nu$ frequency. Both flux densities are measured in mJy.} \par
The indices of about 80\% of the sources from both the LB and LGB samples are smaller than 0.5 and 99\% have indices below 1.0 \citep{U3}. 
The flatness of the radio spectrum can be seen as a consequence of the dominance, also at low radio frequencies, of the inner jet departing from the core relative to the emission radiated by the larger structures of the blazar.  \par
In particular for the  $\gamma$-ray blazar sample, we consider as \textit{A class} candidates the radio sources characterized by $-1.0 \leq \alpha_{352}^{1400}\leq 0.55$ and \textit{B class} as those with $0.55\leq \alpha_{352}^{1400}\leq 0.65$. These two classes have been defined to represent respectively the 80\% (\textit{A class}) and 90\% (\textit{B class}) of $\gamma$-ray blazar radio spectral indices. In fact, the flatness of radio spectrum above 1\,GHz is indeed a well known feature and was already used to discern  $\gamma$-ray blazar associations in the past \citep{CGRABS}. The low-frequency radio observations allow us to confirm this flatness down to 325\,MHz.\par

\subsection{The WISH survey}

 The WISH survey is the natural extension of the WENSS survey  to 1.60 sr of the southern sky. at 352\,MHz. Both surveys were performed in the 1990s by the Westerbork Synthesis Radio Telescope (WSRT), using the bandwidth synthesis mosaicing technique to combine 8 different bands of 5 MHz. WISH covers the area between $-9^\circ < {\rm Dec} < -26^{\circ}$ to a limiting flux density of $\sim 18$ mJy ($5 \sigma$), the same limiting flux density as the WENSS. Due to the low elevation of the observations,  the survey has a lower resolution in declination ($54^{\prime\prime} \times \csc(\delta)$) than in right ascension ($54^{\prime\prime}$), resulting poorer source localization than for WENSS by a factor $\sim 2\times$ on average. Aside from this consideration, the WISH shares with the WENSS the same features upon which this method was calibrated, except for a negligible difference in frequency ($\Delta \nu = 27$ MHz). For this reason, we are confident in applying the same association procedure that we used for WENSS. \par

\subsection{Association procedure}

 For each UGS in the WISH footprint, we search for low-frequency sources in a circular region of radius equal to the semi-major axis of the 95\% confidence level positional uncertainty ellipse of the UGS itself. 
For each WISH source found, we look for a corresponding NVSS source in a circular region of radius equal to 8.5\arcsec\ and we then calculate the radio spectral index $\alpha_{352}^{1400}$.
 For this study, we only consider WISH sources classified as single component sources and featuring a NVSS association.

In addition, we made local maps of the search regions for each UGS, overlaying on the WISH background the brightness contours of the NVSS map, the \textit{Fermi} positional uncertainty ellipse, and any possible blazar-like \textit{WISE} detection up to 3.3\arcsec\ from the position of the matching NVSS source  (see Fig.~\ref{fig:Map2}). This gave us, in addition to a qualitative comparison of the relative positions of various possible candidates and the UGS, a clue about their potential non-blazar nature if complex structures are visible. Also, the angular separation between a candidate and the center of the \fer\ ellipse was taken in account in cases of multiple matches, with the nearest source preferred. \par

For associations between WISH and NVSS and between NVSS and \textit{WISE}, the search radii (of 8.5\arcsec, and 3.3\arcsec, respectively) were selected based on a statistical study, as described in \paperiii.

\subsection{Multi-frequency data}

Finally, we looked for additional multifrequency data for our new blazar candidates and those found in \paperiii. In particular, we searched for matches with sources in the Australia Telescope 20 GHz survey\citep[AT20G,][]{AT20G} for the WISH candidates, and in the Green Bank 4.85 GHz northern sky survey \citep[GB6,][]{GB6} for the WENSS candidates. The AT20G survey, performed between 2005 and 2006 with the Australia Telescope Compact Array (ATCA), contains sources with $S_{20.0}\geq40$ mJy; the GB6, performed between 1986 and 1987 with the NRAO seven-beam receiver, takes into account sources with $S_{4.85}\geq18$ mJy. For these surveys, source association is automatically provided by NASA/IPAC Extragalactic Database (NED). We also refer to \citet{U4} for the analysis of X-ray emission observed by the {\it Swift} X-ray Telescope. \par

\section{Results}

\subsection{UGSs in the WISH footprint}

Of the 575 UGSs in the 2FGL catalog, we discard all that feature a `$c$' analysis flag to rule out  potentially confused or spurious detections due to the strong interstellar emission \citep{2FGL}. The resulting list of UGSs on the WISH footprint has 27 $\gamma$-ray sources. Of these, 17 have at least one WISH and NVSS match (Table~\ref{tab:table}).\par

The total number of candidates is 31. Seven of these candidates have a radio spectral index in the range $\alpha_{325}^{1400}<0.55$  and are considered {\it class A} candidates while six are in the range $0.55 < \alpha_{325}^{1400} < 0.65$ and are considered {\it class B} candidates.

Sixteen out of the 31 candidates in the WISH footprint feature a \wise\ detection in at least two IR bands; three of them are {\it A class} candidates and two are {\it B class}. However, we do not consider \wise\ detection to be necessary for selection. IR data (see Table ~\ref{tab:WISE}) are taken from the \textit{AllWise} November 2013 release \citet{ALLWISE}.  Flat spectrum radio quasars at high redshift or high synchrotron peaked BZBs could be too faint in the IR to be detected in one or more \wise\ bands; moreover, the relative astrometry of the WISH and NVSS {might be poorer} in this region \citep{U5} in respect to WENSS and NVSS.

All considered, we propose eight UGSs blazar-like associations (i.e., five \textit{A} and two \textit{B} class candidates and one special case for WBN 2255.7$-$1833). In particular, we increase the number of {\it class A} and {\it class B} associations combined with \paperiii to 19 and 8, respectively.

\par

\subsection{Multiwavelength observations and radio flatness of low-frequency selected blazar candidates}

In this work and in \citetalias{U3}, we evaluate the flatness of the radio spectrum according to Eqn.~\ref{eq:index} on the basis of just two non-simultaneous flux density measurements,  at 1.4\,GHz (NVSS) and  at low-frequency (352\,MHz for the WISH or 325\,MHz for the WENSS). For this reason, it is important to find other radio flux density data to better investigate the spectral and variability properties of the candidates. We searched the literature  for other radio surveys performed in the WENSS and WISH footprints; we consider here all the candidates found both in \citetalias{U3} and in the present work.\par
\begin{figure}
\centering
\plottwo{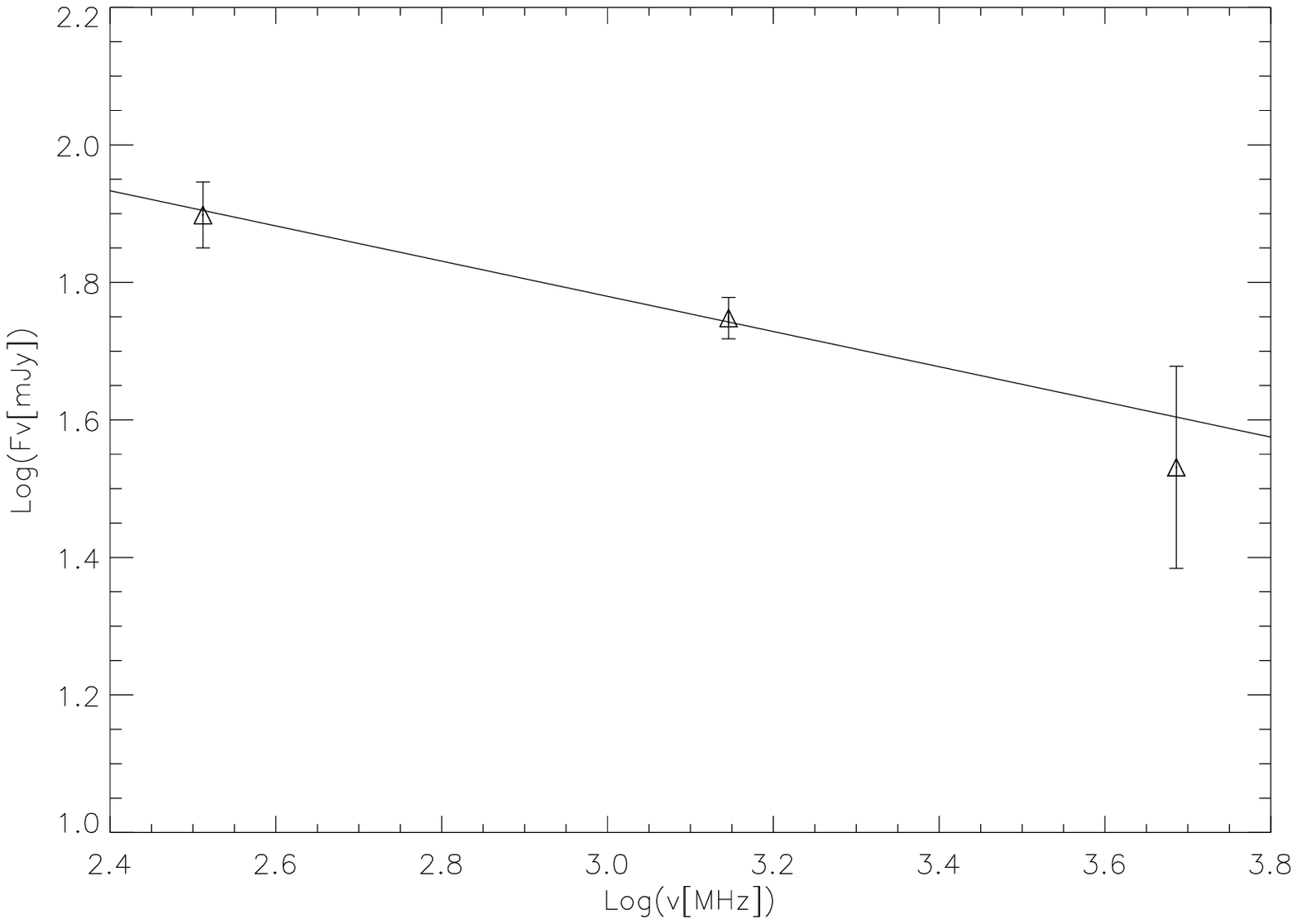}{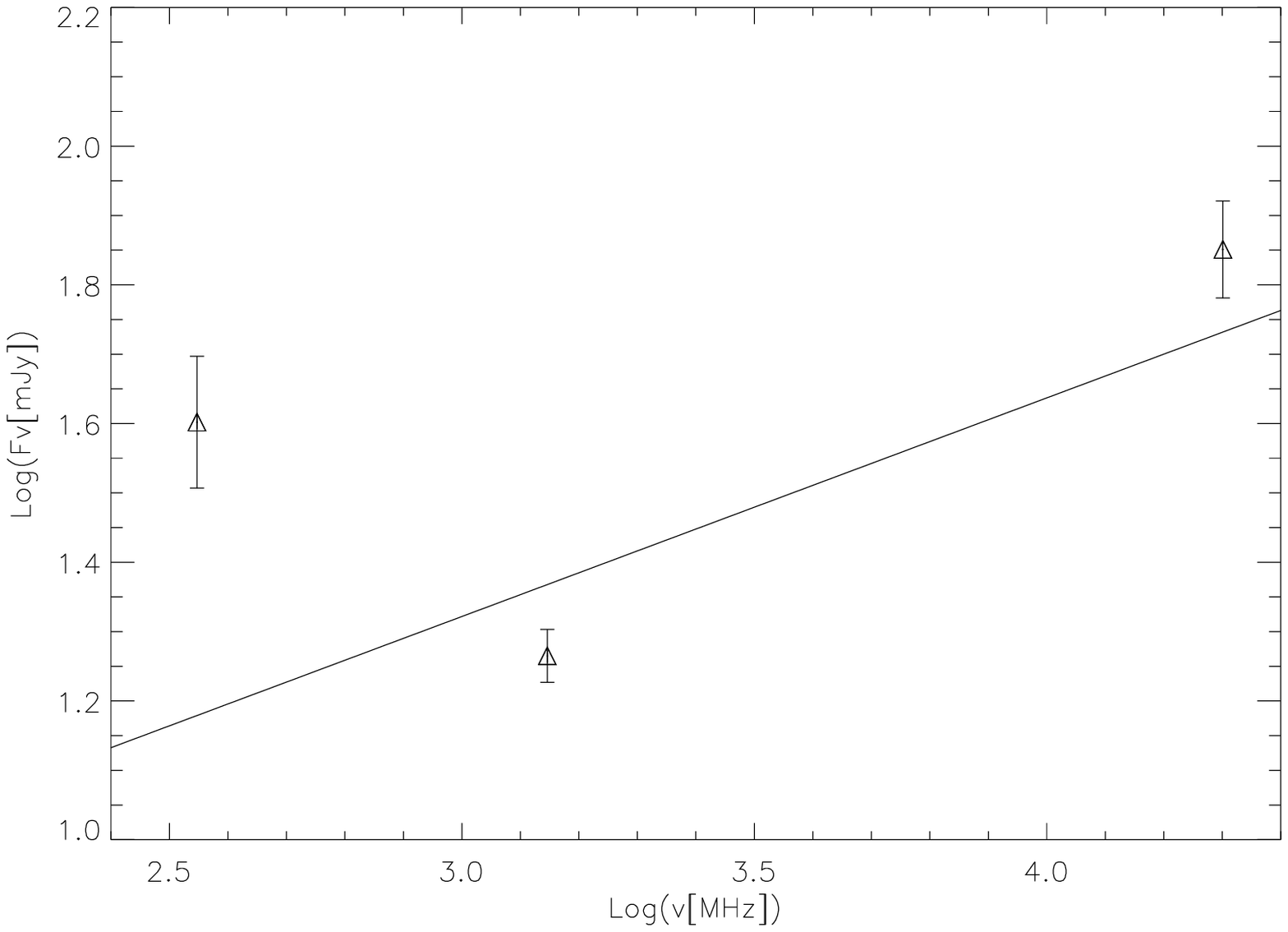}
    \caption{Radio spectra of NVSS J030727+491510 (left panel) and NVSS J125429$-$220419 (right panel). The solid lines represent the linear regression spectral index, which is clearly a good fit to the data only in the first case.}
    \label{fig:regr1}
   \end{figure}

In the WENSS footprint, we found that six \textit{A class} and two \textit{B class} candidates are listed in the GB6 survey at 5\,GHz. We report them in Table \ref{tab:regr1} along with the radio spectral index, obtained with a weighted linear regression on the measurements at the three frequencies. For all these sources, the 5 GHz flux density is in good agreement with the extrapolation of the low-frequency spectrum, confirming that the radio spectrum is flat and flux density variability is not dramatic. A sample spectrum is shown in Fig.~\ref{fig:regr1} (left panel).\par

In the WISH footprint, we find that one \textit{B class} (WNB 1251.8$-$2148, alias NVSS 125429$-$220419) is also listed in the AT20G. Data and radio spectral index regression are reported in Table \ref{tab:regr2} (see also Fig.~\ref{fig:regr1}, right panel). In this case, the extrapolation of the low-frequency power law clearly fails to match the high frequency data, suggesting prominent variability or a rather complex spectral shape, with a strongly inverted component above a few GHz; either way, the source behavior is consistent with the BZB class.\par

\section{Discussion and conclusions}

\subsection{Comparison with Paper III}

As we already noted, the WISH and WENSS surveys are very similar but, a posteriori, we can draw some useful  conclusions regarding the impacts their different features had on our study.
The poorer resolution in declination of WISH ($\sim2\times$ on average) mainly affects the spatial association and not the parameters on which the counterpart is selected. Since spatial association is calibrated with a higher resolution survey like WENSS, applying it to the WISH region and catalog results in a more conservative approach. The fact that we found almost the same ratio (i.e. $\sim1.8$) of candidates per UGS in both WENSS and WISH (58 WENSS candidates for 32 UGSs and 31 WISH candidates for 17 UGSs) indicates that, on average, spatial association is not compromised. \par
Similarly, the resolution discrepancy cannot be blamed for any impact on the quality of the associations proposed among the candidates, given the similar rate of $\gamma$-ray blazar-like associations in the two surveys. In fact, in \citetalias{U3} we propose 21 (including one special association that is neither {\it class A} nor {\it class B}; see  \paperiii) new $\gamma$-ray blazars out of 65 UGSs in the WENSS region (i.e., \ $32\%$) and 8 out of 27 in WISH region (i.e., \ $30\%$)\footnote{The small discrepancy { [with respect to what?]} can be explained, in addition to simple statistics, on the grounds of the slightly different starting list, since in \paperiii\ we considered only $\gamma$-ray sources without {\it any} analysis flag, while here we excluded only the sources with the {\it c} $\gamma$-ray analysis flag in the 2FGL \citep{2FGL}.}. \par

\subsection{Comparison with other methods}

Other methods have been suggested to recognize low-energy counterparts to UGSs, or at least to provide a statistically significant classification of these sources. For instance, \citet{Ackermann2012} have developed a statistical approach to classify UGSs in the first catalog of \fer\ sources \citep[1FGL,][]{Abdo2010}. Six of the 2FGL UGSs in the WISH footprint are associated with 1FGL sources analyzed by \citet{Ackermann2012}, five of which are AGN-like and one of which is pulsar-like. Within the former sample, there are two sources for which we propose an association with a blazar counterpart on the basis of the low-frequency spectrum: 2FGL J2017.5$-$1618 and 2FGL J2358.4$-$1811. Interestingly, for the single UGS for which \citet{Ackermann2012} propose a pulsar classification (2FGL J1544.5$-$1126), our method finds a WENSS-NVSS match with quite steep spectral index ($\alpha=0.74\pm0.03$), rejecting a blazar scenario.

\begin{figure}
\centering
\plotone{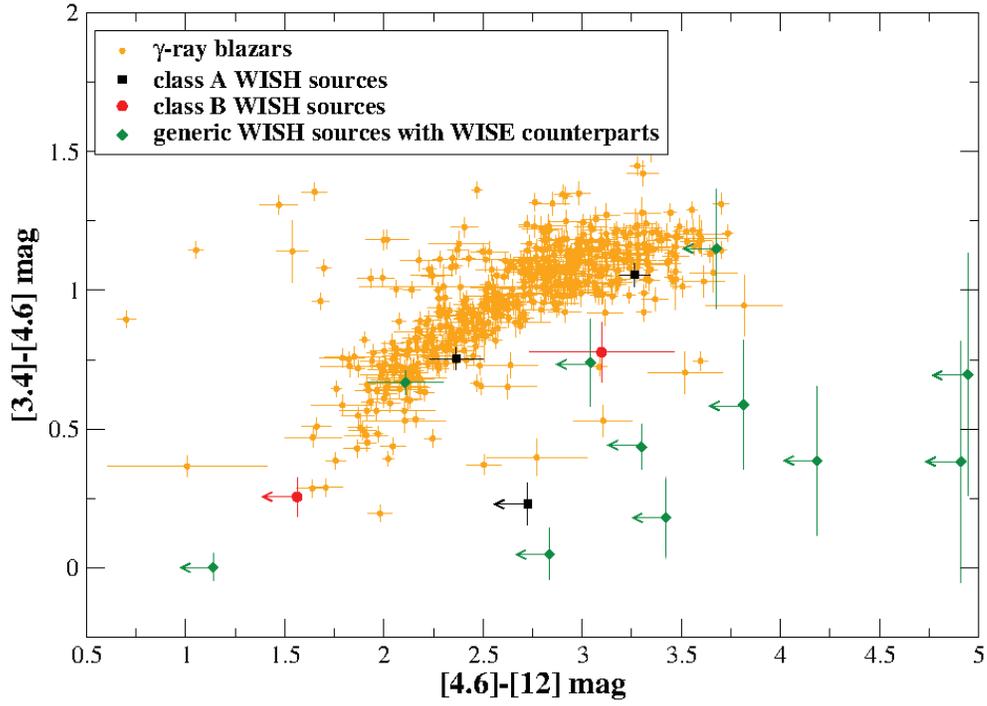}
    \caption{
The $[3.4]-[4.6]-[12] \, \mu$m color-color plot of all the \wise\ counterparts of the WISH-NVSS sources (green diamonds) and candidate blazars (black squares:\ {\it class A}; red squares:\ {\it class B}) in comparison with the blazars that constitute the WISE $\gamma$-ray strip (orange dots). }
    \label{f.wgs}
   \end{figure}

We further compared our results with the proposed associations found in the other papers of this series. In Figure~\ref{f.wgs} we show the comparison between the distribution in the IR  $[3.4]-[4.6]-[12]\, \mu$m color-color plane provided by \wise\ for the $\gamma$-ray emitting blazars and the sources selected in this work. The overall distribution of the whole set of the simultaneous WISH-NVSS matches (black, red and green dots) is quite more scattered than the $\gamma$-ray blazar population (orange dots). However, when we only consider the low-frequency selected blazar candidates, all of the most prominent outliers are excluded and the {remaining five sources} (black and red dots) are in much better agreement with the IR colors of $\gamma$-ray blazars. In two cases, the $A$-class candidates NVSS 120900$-$231335 and NVSS 222830$-$163643, the agreement is very good. For the latter source, associated with 2FGL\, J2228.6$-$1633,  our method also provides the same counterpart candidate selected on the basis of the kernel density estimator technique to IR colors of WISE counterparts applied to X-ray \citep{U4} and radio \citep{U5} data.

\subsection{Radio flux density analysis}

The $\gamma$-ray blazar candidates selected in this work and in \paperiii\ have by selection the same radio spectral properties of confirmed Roma-BZCAT blazars and of 2FGL blazar associations. It is natural to wonder why these sources have not been detected in the 2FGL catalog and eventually associated, e.g.\ in the second catalog of AGNs detected by \fer\ \citep{2LAC}. The sources could have been excluded from the 2LAC because they do not formally pass the threshold for being considered high confidence associations (a test that does not take into account the spectral index); typically, this could happen for low flux density radio sources,  which have a larger spatial density. It is thus likely that, together with a general similarity to the already known blazars, our candidates also have some peculiar characteristics.  

\begin{figure*}
\centering
\plottwo{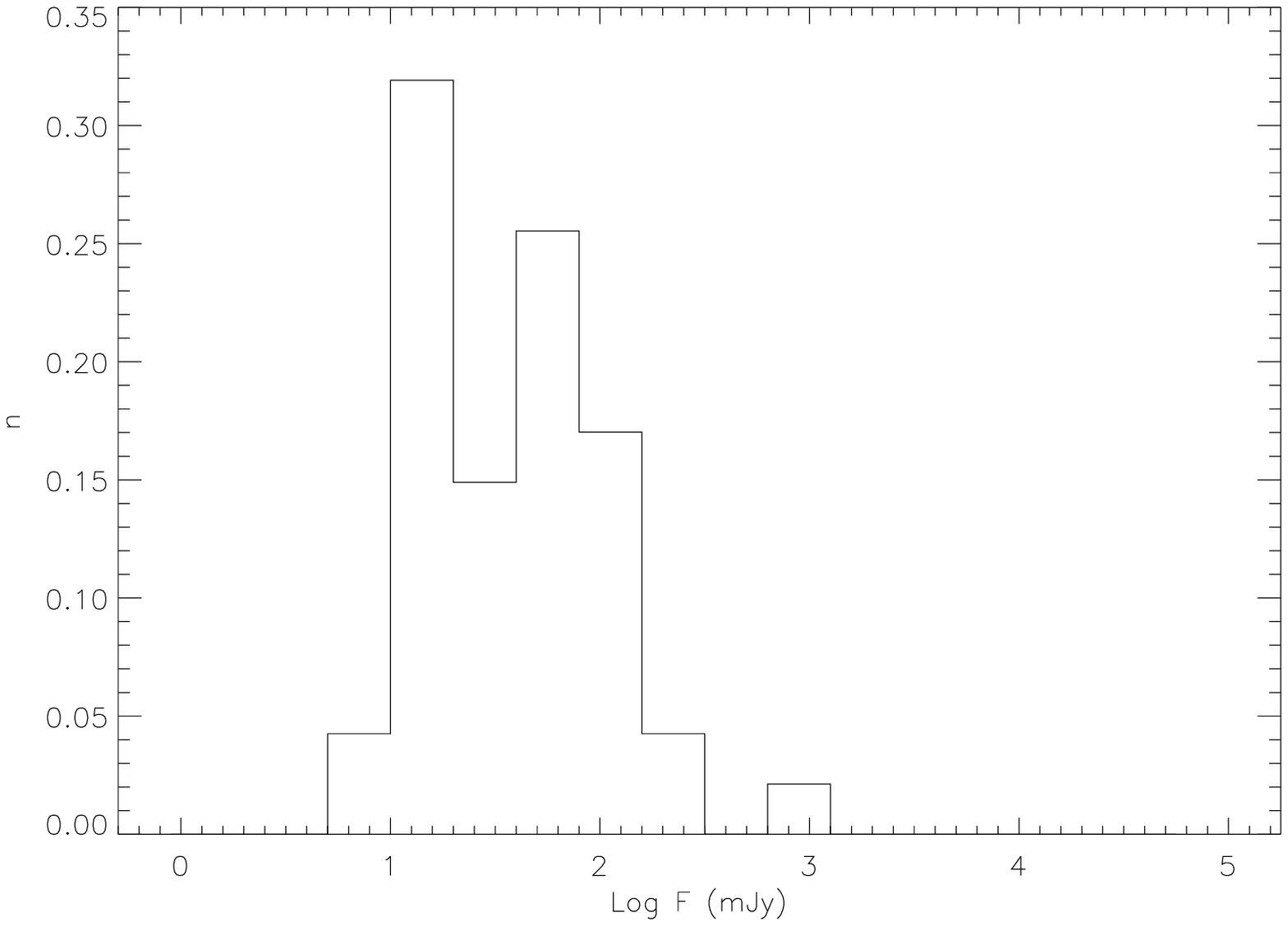}{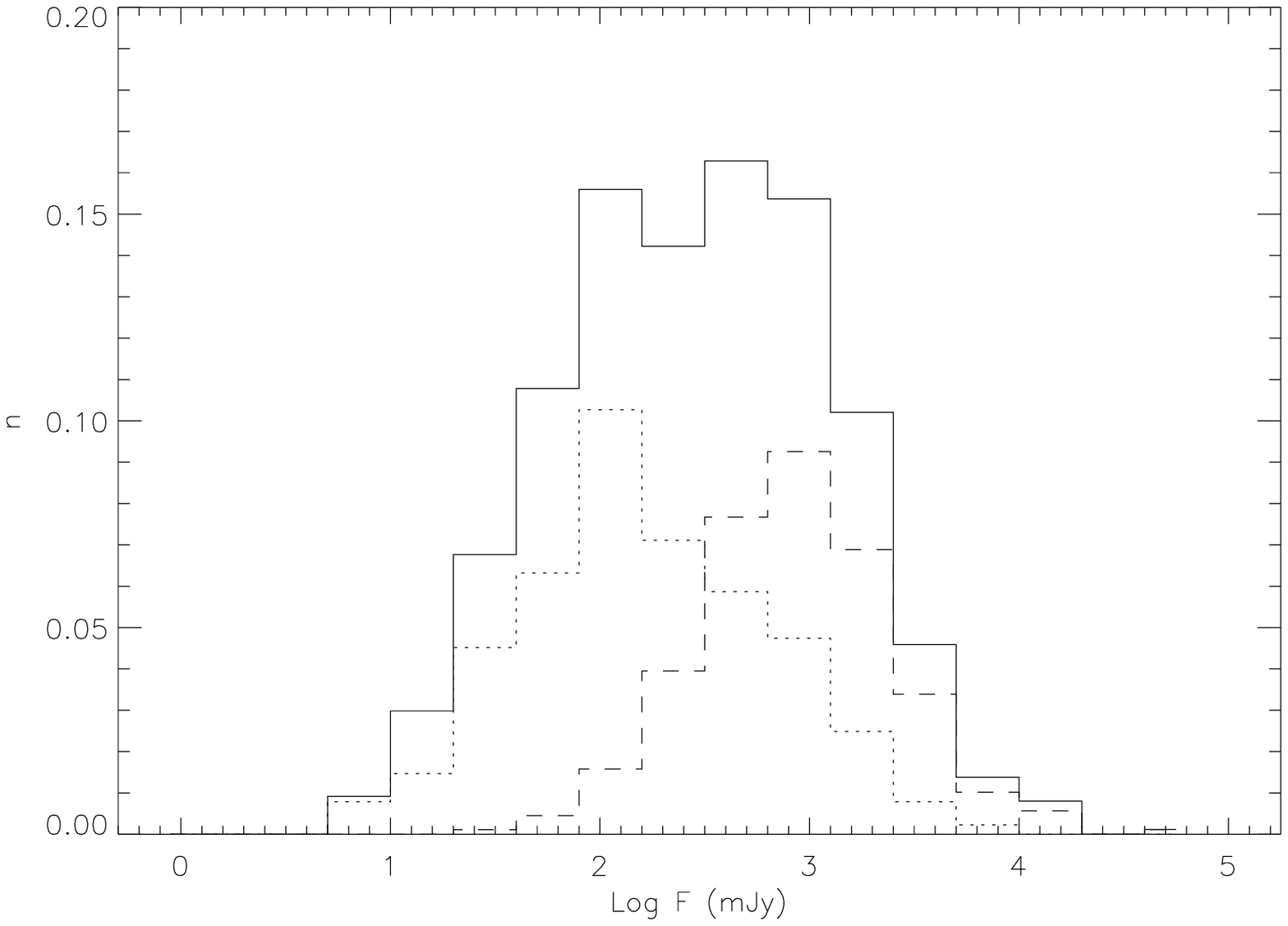}
    \caption{Normalized distributions of flux density at 1.4\,GHz. Left panel: $\gamma$-ray blazar candidates selected among the UGSs in the WENSS and WISH footprints; right panel: sources in the 2LAC (solid line: all 2LAC sources, dotted line: BZBs, dashed line: BZQs).}
    \label{fig:hist}
   \end{figure*}

For this reason, we show in Fig.~\ref{fig:hist} the distribution of radio flux density at 1.4\,GHz for all the blazars in the 2LAC and for the candidates selected here and in \paperiii. The two distributions are clearly different, as confirmed by a Kolmogorov–Smirnov test which yields a probability of $2.7\times10^{-15}$ of being obtained from the same population. In particular, the 2LAC blazar flux density distribution is shifted to much larger values. This strongly suggests that our method is very efficient at selecting faint blazars. These sources are potentially of great interest: if they are of the BZB type, it could mean that they could be of the extreme and elusive class of ultra-high synchrotron peaked sources; one prominent example could be WNB 2225.8$-$1652, which has an inverted  low-frequency radio spectral index of $\alpha=-0.19\pm0.10$ and is our proposed association for the UGS 2FGL\, J2228.6$-$1633, characterized by a $\gamma$-ray spectrum as hard as $\Gamma=2.07\pm0.16$. On the other hand, some of our candidates could also be faint BZQs, and in this case their low flux densities could stem from their high redshifts \citep[like WN 1500.0+4815 with an estimated redshift of 2.78, see][]{U3}, which would also make them of scientific value. No WISH sources have a measured redshift.

\subsection{Summary and outlook}

We have searched for counterparts of the the 27 UGSs in the 1.6 sr footprint of the WISH survey, following the methods described in \paperiii\ and based on the flat spectrum at low-frequency characteristic of blazars. We have found blazar-like associations for 8 UGSs, that together with the 23 sources selected in the WENSS footprint with the same method provides a blazar association for a sample of 30 new $\gamma$-ray blazar associations. This sample extends the distribution of the radio flux density of $\gamma$-ray blazars to lower values, allowing us to study otherwise elusive AGNs.

The application of our method thus shows promising results in terms of numbers of counterparts proposed and their physical  features. In particular, the possibility of using low-frequency radio data to find UGSs counterparts is even more important in the light of the  imminent start of numerous studies in this frequency range, like the LOw Frequency ARray \citep[LOFAR,][]{van Haarlem2013},  the Murchison Widefield Array \citep[MWA,][]{Tingay2013}, the Long Wavelength Array \citep[LWA,][]{Ellingson2009}, and eventually the Square Kilometer Array \citep[SKA, e.g.][]{Dewdney2010}. These facilities will allow our method to be extended using an even deeper and simultaneous dataset, while dedicated targeting of the candidates with optical spectroscopy and Very Long Baseline Interferometry observations will confirm the natures of the proposed associations. \par

\acknowledgements 
We are grateful to S. Digel for providing us useful suggestions that improved the presentation of our results. 
M.\ Giroletti acknowledges financial contribution from grant PRIN-INAF-2011. The work is supported by the NASA grants NNX12AO97G and NNX13AP20G.
R.\ D'Abrusco gratefully acknowledges the financial support of the US Virtual Astronomical Observatory, which is sponsored by the National Science Foundation and the National Aeronautics and Space Administration.
The work by G.\ Tosti is supported by the ASI/INAF contract I/005/12/0.
This research has made use of data obtained from the High Energy Astrophysics Science Archive Research Center (HEASARC) provided by NASA's Goddard Space Flight Center; the SIMBAD database operated at CDS, Strasbourg, France; the NASA/IPAC Extragalactic Database (NED) operated by the Jet Propulsion Laboratory, California Institute of Technology, under contract with the National Aeronautics and Space Administration.  Part of this work is based on the NVSS (NRAO VLA Sky Survey); The National Radio Astronomy Observatory is operated by Associated Universities, Inc., under contract with the National Science Foundation.  This publication makes use of data products from the Wide-field Infrared Survey Explorer, which is a joint project of the University of California, Los Angeles, and the Jet Propulsion Laboratory/California Institute of Technology, funded by the National Aeronautics and Space Administration.

\appendix

\subsection{Notes on individual UGSs}

Below, we report a brief analysis of the results for each UGS listed in Table~\ref{tab:table}.

\begin{figure*}
\centering
\plottwo{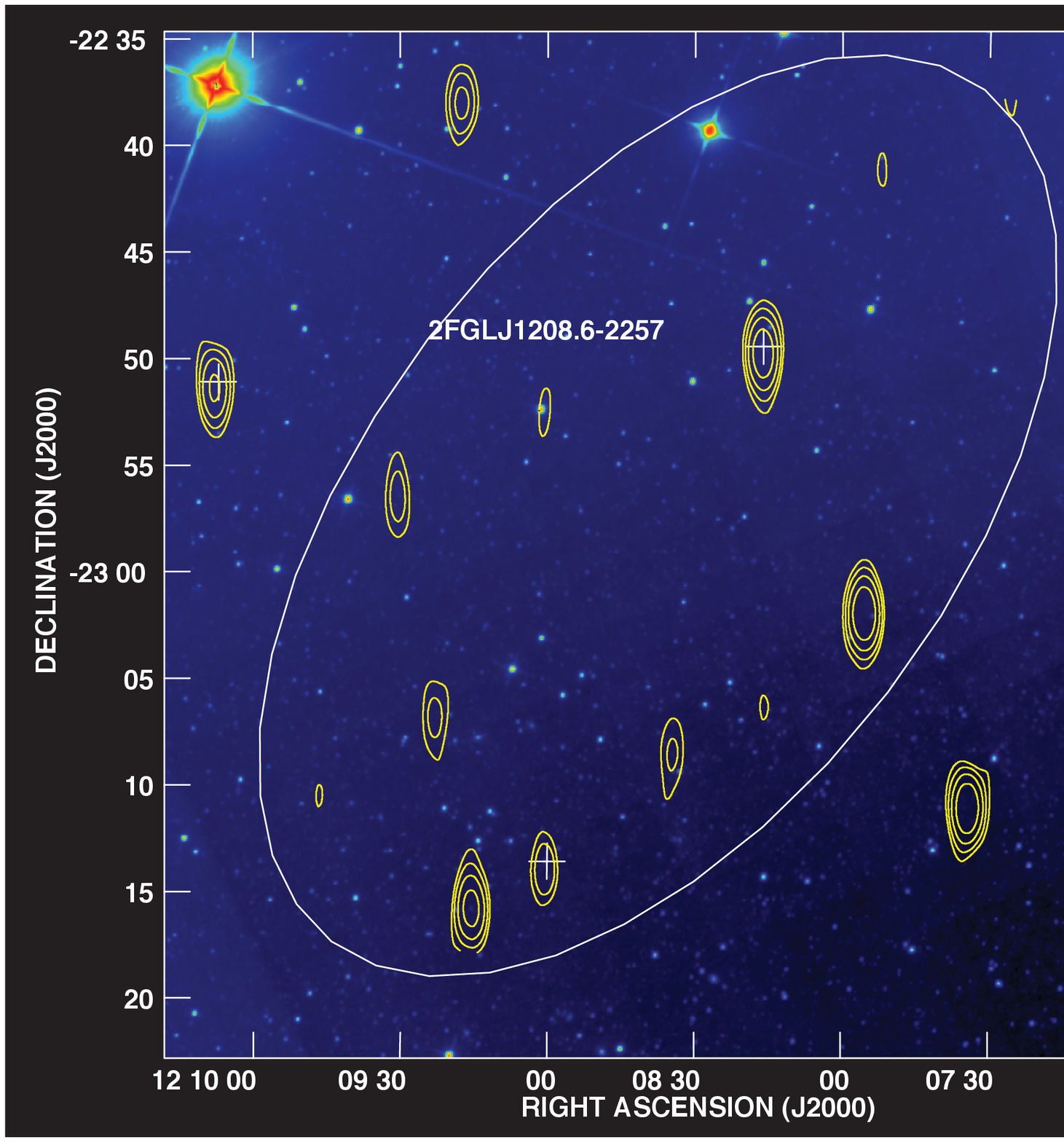}{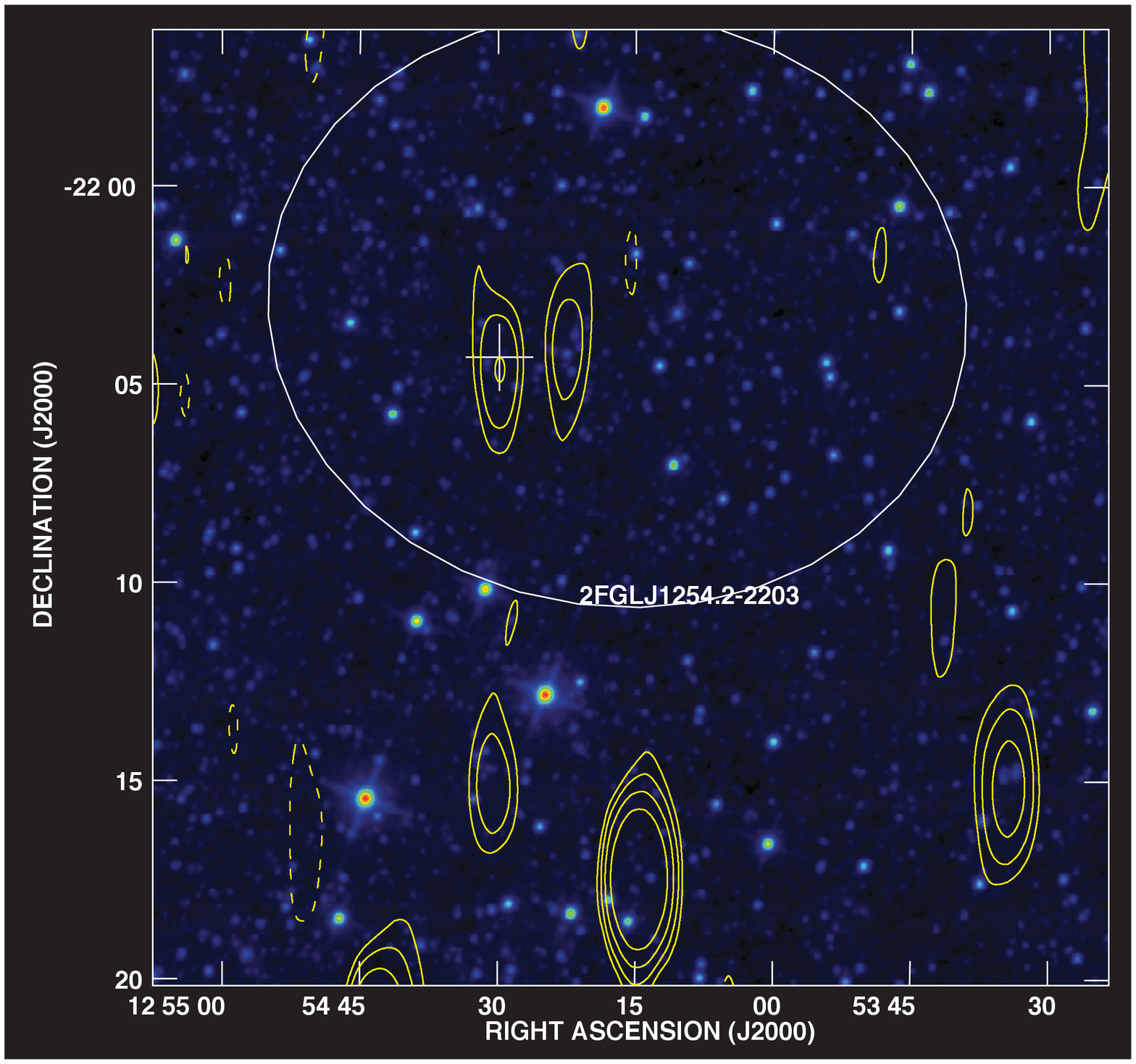}
    \caption{Map of the search regions around 2FGL J1208$-$2257 (left panel) and 2FGL J1254.2$-$2203 (right panel). Color scale background shows the WISE image, contours represent the 352 MHz emission from the WISH, and crosses indicate NVSS sources. The white ellipse is the $\gamma$-ray 95\% confidence region. In the right-hnad panel, our candidate is the leftmost of the twin NVSS sources.  (Note that the righthand source is formally not in the WISH catalog.)}
    \label{fig:Map2}
   \end{figure*}

\begin{itemize}
\item {\it 2FGL J0340.7$-$2421} Both candidates in this search region have an IR detection by \wise. Our method selects WN 0338.4$-$2436 as a blazar association due to its flatter ({\it class A}) spectral index; this source also has a greater flux density.
\item	 {\it 2FGL J0600.8$-$1949} There are three candidates: WN 0558.8$-$1950 provides the best association among the others since it is nearest to the $\gamma$-ray position 
and it has blazar-like IR colors, a flatter spectrum, and larger flux density; however, the radio spectrum is steeper than 0.71 so we can not formally classify it as a blazar.
\item {\it 2FGL J1059.9$-$2051}	Both candidates in the search region are detected in the IR by \wise and have large flux density but have a quite steep radio spectral index.
\item {\it 2FGL J1208.6$-$2257}
	WN 1206.4$-$2256 is the best association among the other candidates because of its flat ({\it class A}) spectral index and \wise\ blazar-like detection.
\item  {\it 2FGL J1254.2$-$2203} WNB 1251.8$-$2148 has an intermediate spectral index and is proposed as a {\it class B} blazar association. We note that a second nearby 1.4 GHz source (NVSS 125422-220413, see Fig.~\ref{fig:Map2}, right panel) is not present in the WISH survey, suggesting a flatter spectral index; this source could also be a blazar-like candidate and a partial contributor to the $\gamma$-ray emission in this region.
\item	 {\it 2FGL J1458.5$-$2121} The only candidate has the IR features of a blazar-like source but a too-steep spectral index; moreover, a  much brighter source that could be the association for the UGS lies just outside the search region. 
\item {\it 2FGL J1544.5$-$1126}	The only candidate has a too-steep radio spectrum and no proper \wise\ match, so it is not a likely blazar association. Indeed, this UGS was already detected in the 1FGL and the statistical method of \citet{Ackermann2012} suggested that it should be classified as a pulsar.
\item {\it 2FGL J1624.2$-$2124} Four candidates are present within the $\gamma$-ray error radius, some with {\it A} or {\it B class} spectral index, and one with a \wise\ detection. The most promising potential associations are WNB 1620.7$-$2120 (flattest $\alpha$) and WNB 1621.1$-$2119 (largest flux density and presence of IR emission); we prefer the first, consistent with the method.  
\item {\it 2FGL J1631.0$-$1050} None of the three candidates within the search region has a flat spectral index. 
\item {\it 2FGL J1646.7$-$1333}
	WN 1644.0$-$1323, even with a \wise\ detection and typical blazar-like X emission \citep{U4}, has a too-steep radio spectral index to be considered a blazar-like association.
\item {\it 2FGL J1913.8$-$1237}
	According to our method, none of the candidates in this region can be considered a good blazar association.
\item {\it 2FGL J2009.2$-$1505}
	The only candidate, even featuring blazar-like X emission \citep{U4}, has a somewhat too steep spectral index for being considered a blazar-like source, and it also lacks a \wise\ detection. 
\item {\it 2FGL J2017.5$-$1618} WNB 2014.9$-$1627 is a {\it class A} blazar association, even if it would not be selected as such on the basis of the IR emission. \citet{Ackermann2012} classify this UGS as a likely AGN.
\item {\it 2FGL J2031.4$-$1842}	Both candidates in the search region have a {\it class B}  radio spectral index; the most likely blazar association is WNB 2027.8$-$1900, which is brighter and detected in the IR.
\item {\it 2FGL J2124.0$-$1513}		The only candidate has a too-steep spectral index for being selected as a blazar-like source, even though it has a \wise\ detection. 
\item {\it 2FGL J2228.6$-$1633}	The inverted spectral index of WNB 2225.8$-$1652, its blazar-like \wise\ colors, and X-ray emission \citep{U4} make this source a highly reliable blazar association for the UGS. 
\item {\it 2FGL J2358.4$-$1811}	WNB 2355.7$-$1833 has a radio spectral index just above our {\it B class} threshold and a \wise\ detection.  Although its IR colors are not very blazar-like, the error range on the spectral index overlaps with the {\it B class} range. We consider it a suitable, even if not formal, blazar-like association, in agreement with the AGN classification proposed for this UGS by \citet{Ackermann2012}.
\end{itemize}

\begin{deluxetable}{c c c l l c c c}
\tabletypesize{\scriptsize}
\tablewidth{0pt}
\tablecaption{Possible blazar-like counterparts for UGSs in the WISH region. For each source, we list the 2FGL name (Col.\ 1), the list of candidates per UGS with WISH name (Col.\ 2) and NVSS name (Col.\ 3), the corresponding flux density at 352 MHz (Col.\ 4) and 1400 MHz (Col.\ 5), the presence in the \wise\ catalog (Col.\ 6), the low-frequency radio spectral index (Col.\ 7) and the relative classification (Col.\ 8). \label{tab:table} }
\tablehead{
\colhead{2FGL name} & \colhead{WISH name}  & \colhead{NVSS name}  & \colhead{$S_{352}$} & \colhead{$S_{1400}$} & \colhead{\textit{WISE}}  &  \colhead{$\alpha^{1400}_{352}$}  & \colhead{Index} \\
\colhead{} & \colhead{} & \colhead{} & \colhead{(mJy)} & \colhead{(mJy)} & \colhead{detection} & \colhead{} & \colhead{Class}  }
\startdata
2FGL J0340.7$-$2421 & WNB 0338.0$-$2425 & NVSS 034011$-$241602 & $72\pm4$  & $23.5\pm1.1$ & $\surd$ & $0.81\pm0.04$ &  \\ 
 		  & WNB 0338.4$-$2436 & NVSS 034033$-$242712 & $111\pm4$ & $82\pm3$     & $\surd$ & $0.22\pm0.03$ &  A\\ 
2FGL J0600.8$-$1949 & WNB 0557.9$-$1954 & NVSS 060003$-$195446 & $44\pm4$  & $12.7\pm0.6$ & $\surd$ & $0.90\pm0.06$ &  \\ 
     		  & WNB 0558.8$-$1950 & NVSS 060100$-$195049 & $256\pm4$ & $96\pm3$     & $\surd$ & $0.71\pm0.02$ &  \\ 
     		  & WNB 0559.4$-$1948 & NVSS 060138$-$194853 & $124\pm4$ & $43.2\pm1.4$ &         & $0.76\pm0.03$ &  \\ 
2FGL J1059.9$-$2051 & WNB 1057.1$-$2037 & NVSS 105935$-$205311 & $414\pm8$ & $138\pm5$    &  $\surd$      & $0.80\pm0.02$ &  \\ 
     		  & WNB 1057.7$-$2040 & NVSS 110014$-$205621 & $737\pm4$ & $232\pm7$    &   $\surd$      & $0.84\pm0.02$ &  \\ 
2FGL J1208.6$-$2257 & WNB 1205.6$-$2232 & NVSS 120816$-$224925 & $151\pm3$ & $66\pm2$	    & 	      & $0.60\pm0.02$ &  B\\ 
   		  & WNB 1206.4$-$2256 & NVSS 120900$-$231335 & $34\pm3$  & $28.8\pm1.0$ & $\surd$      & $0.12\pm0.06$ &  A\\ 
   		  & WNB 1207.5$-$2234 & NVSS 121007$-$225106 & $96\pm3$  & $35\pm1.1$   & 	      & $0.73\pm0.03$ &  \\ 
2FGL J1254.2$-$2203 & WNB 1251.8$-$2148 & NVSS 125429$-$220419 & $40\pm4$  & $18.4\pm0.7$ & 	      & $0.56\pm0.06$ &  B\\ 
2FGL J1458.5$-$2121 & WNB 1456.2$-$2112 & NVSS 145904$-$212357 & $190\pm3$ & $70\pm2$     & $\surd$ & $0.73\pm0.02$ &  \\ 
2FGL J1544.5$-$1126 & WNB 1541.4$-$1115 & NVSS 154414$-$112443 & $83\pm3$  & $30.0\pm1.0$ &	      & $0.74\pm0.03$ &  \\ 
2FGL J1624.2$-$2124 & WNB 1620.5$-$2058 & NVSS 162332$-$210457 & $19\pm3$  & $9.9\pm0.6$  &	      & $0.47\pm0.10$ &  A\\ 
   		  & WNB 1620.7$-$2120 & NVSS 162345$-$212716 & $30\pm3$  & $19.2\pm1.5$ & 	      & $0.32\pm0.07$ &  A\\ 
   		  & WNB 1621.1$-$2119 & NVSS 162403$-$212645 & $175\pm3$ & $80\pm3$    & $\surd$      & $0.57\pm0.02$ &  B\\ 
   		  & WNB 1623.0$-$2111 & NVSS 162600$-$211825 & $105\pm3$ & $45.2\pm1.4$ & 	      & $0.61\pm0.03$ &  B\\ 
2FGL J1631.0$-$1050 & WNB 1628.0$-$1045 & NVSS 163049$-$105218 & $106\pm5$ & $29.7\pm1.0$ & $\surd$ & $0.92\pm0.03$ &  \\ 
   		  & WNB 1628.7$-$1037 & NVSS 163130$-$104322 & $129\pm5$ & $45.6\pm1.8$ & 	      & $0.75\pm0.03$ &  \\ 
   		  & WNB 1628.8$-$1044 & NVSS 163139$-$105057 & $591\pm5$ & $184\pm6$    & $\surd$ & $0.85\pm0.02$ &  \\ 
2FGL J1646.7$-$1333 & WNB 1644.0$-$1323 & NVSS 164651$-$132849 & $447\pm4$ & $99\pm3$     & $\surd$ & $1.09\pm0.02$ &  \\ 
2FGL J1913.8$-$1237 & WNB 1910.5$-$1235 & NVSS 191320$-$122949 & $44\pm4$  & $14.7\pm0.6$ & 	      & $0.79\pm0.06$ &  \\ 
   		  & WNB 1910.8$-$1246 & NVSS 191339$-$124120 & $133\pm4$ & $39.7\pm1.3$ & 	      & $0.88\pm0.03$ &  \\ 
2FGL J2009.2$-$1505 & WNB 2005.8$-$1513 & NVSS 200838$-$150500 & $299\pm6$ & $105\pm4$    & 	      & $0.76\pm0.02$ &  \\ 
2FGL J2017.5$-$1618 & WNB 2014.9$-$1627 & NVSS 201745$-$161820 & $110\pm3$ & $53.6\pm1.7$ & 	      & $0.52\pm0.02$ &  A\\ 
2FGL J2031.4$-$1842 & WNB 2027.6$-$1850 & NVSS 203030$-$184033 & $204\pm3$ & $92\pm3$	    & 	      & $0.58\pm0.02$ &  B\\ 
   		  & WNB 2027.8$-$1900 & NVSS 203044$-$185033 & $324\pm3$ & $134\pm4$    & $\surd$      & $0.64\pm0.02$ &  B\\ 
2FGL J2124.0$-$1513 & WNB 2121.8$-$1533 & NVSS 212438$-$152017 & $147\pm3$ & $40.0\pm1.3$ & $\surd$ & $0.94\pm0.02$ &  \\ 
2FGL J2228.6$-$1633 & WNB 2225.8$-$1652 & NVSS 222830$-$163643 & $16\pm3$  & $20.9\pm1.1$ & $\surd$ & $-0.19\pm0.10$ &  A\\ 
   		  & WNB 2226.0$-$1641 & NVSS 222842$-$162619 & $16\pm3$  & $7.2\pm0.5$  & 	      & $0.58\pm0.10$ &  A\\ 
2FGL J2358.4$-$1811 & WNB 2355.7$-$1833 & NVSS 235820$-$181621 & $25\pm3$  & $9.7\pm0.6$  & $\surd$ & $0.69\pm0.09$ &  
\enddata
\end{deluxetable}

\begin{deluxetable}{c c c c c}
\tabletypesize{\scriptsize}
\tablewidth{0pt}
\tablecaption{\textit{WISE} [3.4]-[4.6] and [4.6]-[12] magnitudes. Data are expressed in terms of magnitudes (Vega system) and represent the difference between total in-band brightness measurements. The [4.6]-[12] error is not present whenever the [12] band detection is an upper limit. \label{tab:WISE}}
\tablehead{
\colhead{NVSS name} & \colhead{[3.4]-[4.6] mag} & \colhead{[3.4]-[4.6] error mag} & \colhead{[4.6]-[12] mag} & \colhead{[4.6]-[12] error mag} }
\startdata
  NVSS 034011-241602 & 0.59 & 0.23 & 3.81 & \\
  NVSS 034033-242712 & 0.23 & 0.08 & 2.73 & \\
  NVSS 060003-195446 & 0.44 & 0.08 & 3.30 & \\
  NVSS 060100-195049 & 0.05 & 0.09 & 2.83 & \\
  NVSS 105935-205311 & 0.74 & 0.16 & 3.04 & \\
  NVSS 110014-205621 & 0.39 & 0.27 & 4.19 & \\
  NVSS 120900-231335 & 1.06 & 0.04 & 3.27 & 0.08 \\
  NVSS 145904-212357 & 1.15 & 0.22 & 3.69 & \\
  NVSS 162403-212645 & 0.26 & 0.07 & 1.56 & \\
  NVSS 163049-105218 & 0.70 & 0.44 & 4.95 & \\
  NVSS 163139-105057 & 0.00 & 0.05 & 1.14 & \\
  NVSS 164651-132849 & 0.67 & 0.04 & 2.11 & 0.19 \\
  NVSS 203044-185033 & 0.78 & 0.11 & 3.01 & 0.38 \\
  NVSS 212438-152017 & 0.38 & 0.44 & 4.91 & \\
  NVSS 222830-163643 & 0.75 & 0.04 & 2.37 & 0.14 \\
  NVSS 235820-181621 & 0.18 & 0.15 & 3.42 & \\
\enddata
\end{deluxetable} 

\begin{deluxetable}{c c c c c c c}
\tabletypesize{\scriptsize}
\tablewidth{0pt}
\tablecaption{Regression of radio spectral index with GB6 data of WENSS possible blazar-like counterparts.   \label{tab:regr1}}
\tablehead{
\colhead{NVSS name}  & \colhead{Class} & \colhead{$\log{S_{325}}$ [mJy]} & \colhead{$\log{S_{1400}}$ [mJy]} &   \colhead{$\log{S_{4850}}$ [mJy]} &  \colhead{$\alpha_{\rm regr}$} & \colhead{$\alpha^{1400}_{325}$} }
\startdata
NVSS J030727+491510 & A & $ 1.90 \pm 0.05 $ & $ 1.75 \pm 0.03 $ & $ 1.53 \pm 0.15 $ & $ 0.26 \pm 0.08 $ & $ 0.24 \pm 0.04 $ \\
NVSS J033153+630814 & A & $ 1.79 \pm 0.05 $ & $ 1.63 \pm 0.03 $ & $ 1.40 \pm 0.16 $ & $ 0.28 \pm 0.09 $ & $ 0.26 \pm 0.04 $ \\
NVSS J035309+565431 & A & $ 1.98 \pm 0.05 $ & $ 1.76 \pm 0.03 $ & $ 1.64 \pm 0.11 $ & $ 0.32 \pm 0.08 $ & $ 0.34 \pm 0.04 $ \\
NVSS J072354+285930 & A & $ 1.90 \pm 0.06 $ & $ 1.56 \pm 0.03 $ & $ 1.49 \pm 0.16 $ & $ 0.48 \pm 0.10 $ & $ 0.53 \pm 0.04 $ \\
NVSS J150229+555204 & A & $ 1.70 \pm 0.06 $ & $ 1.54 \pm 0.03 $ & $ 1.36 \pm 0.17 $ & $ 0.26 \pm 0.10 $ & $ 0.25 \pm 0.05 $ \\
NVSS J150147+480335 & A & $ 1.43 \pm 0.12 $ & $ 1.32 \pm 0.03 $ & $ 1.41 \pm 0.15 $ & $ 0.06 \pm 0.16 $ & $ 0.18 \pm 0.08 $ \\
NVSS J210805+365526 & A & $ 1.83 \pm 0.05 $ & $ 1.88 \pm 0.03 $ & $ 1.83 \pm 0.10 $ & $ -0.04 \pm 0.08 $ & $ -0.08 \pm 0.04 $ \\
NVSS J060102+383828 & B &  $ 3.261 \pm 0.002 $ & $ 2.85 \pm 0.03 $ & $ 2.51 \pm 0.09 $ & $ 0.65 \pm 0.48 $ & $ 0.65 \pm 0.02 $ \\
NVSS J101657+560112 & B & $ 2.00 \pm 0.04 $ & $ 1.61 \pm 0.03 $ & $ 1.79 \pm 0.10 $ & $ 0.44 \pm 0.06 $ & $ 0.62 \pm 0.03 $
\enddata
\end{deluxetable}

\begin{deluxetable}{c c c c c c c}
\tabletypesize{\scriptsize}
\tablewidth{0pt}
\tablecaption{Regression of radio spectral index with AT20G data of WISH possible blazar-like counterparts.  \label{tab:regr2}}
\tablehead{
\colhead{NVSS name}  & \colhead{Class} & \colhead{$\log{S_{352}}$ [mJy]} & \colhead{$\log{S_{1400}}$ [mJy]} &   \colhead{$\log{S_{20000}}$ [mJy]} &  \colhead{$\alpha_{\rm regr}$} & \colhead{$\alpha^{1400}_{352}$} }
\startdata
NVSS J125429$-$220419 & B & $ 1.60 \pm 0.10 $ & $ 1.26 \pm 0.04 $ & $ 1.85 \pm 0.07 $ & $ -0.31 \pm 0.06 $ & $ 0.56 \pm 0.06 $
\enddata
\end{deluxetable}

\end{document}